\documentclass[epj]{svjour}
\usepackage{latexsym}
\usepackage{graphics}

\begin{document}

\title{A Field-Effect-Transistor from Graphite: No Effect of Low Gate Fields}
\author{H. Kempa \and P. Esquinazi}
\institute{Abteilung Supraleitung und Magnetismus\thanks{e-mail:
sum@physik.uni-leipzig.de}, Universit\"at Leipzig, Linn\'estr. 5,
D-04103 Leipzig, Germany}
\date{Received: date / Revised version: date}
\abstract{
Inspired by the striking similarities between the
metal-insulator transitions in graphite and Si-MOSFET's and the
recent attention to charge doping in carbon-based materials, we
have made attempts to fabricate a field-effect transistor based on
graphite. A relatively thick layer of boron nitride turned out to
be able to serve as a gate dielectric. This, however, limits the
achievable electric gate field, which might be the reason for our
observation of no charge-doping effect.
}
\maketitle

The striking similarities between Si-MOSFET's and highly ordered
graphite have been pointed out in a number of recent publications
\cite{Kempa1,Kempa2,Kempa3,Kopelevich1}. Namely, the essentially
two-dimensional character of the electron system and the basically
equal charge carrier density manifest themselves in
metal-insulator transitions in both systems with remarkable
similar properties. The main difference remains that the
transition in Si-MOSFET's can be driven by the charge carrier
density, which is tuned through charge doping in the field-effect
transistor configuration \cite{Abrahams}, while in graphite the
transition is driven only by a magnetic field applied
perpendicular to the graphite layers \cite{Kempa3}. In
Si-MOSFET's, however, it has been demonstrated that a parallel
field can drive the transition \cite{Simonian}.

Recent reports on charge-doping effects in carbon-based materials
\cite{Schoen1,Schoen2} have attracted great attention but so far
any attempts to reproduce the results failed (see e.g.
\cite{Lee}), supporting the revelation that a major part of those
results was pure invention \cite{Beasley}. Nevertheless it has
been pointed out, that it's not unlikely to observe such effects
and some ways have been proposed \cite{Ball}.

Clearly, it is desirable to fabricate a field-effect transistor
based on graphite, like any research that is potentially able to
deepen the similarities between graphite and Si-MOSFET's and thus
shed light on both of the systems. In this letter we report on our
attempts to do so.

Our sample is highly oriented pyrolytic graphite (HOPG) produced
by the research institute "Graphite" in Moscow with dimensions
$4.7 \times 2.6 \times 0.72 \mbox{ mm}^3$. X-ray diffraction
measurements give the crystal lattice parameters $a = 2.48 \mbox{
\AA}$ and $c = 6.71 \mbox{ \AA}$, the density is $2.26 \mbox{
g}\cdot\mbox{cm}^{-3}$. The high degree of orientation of the
crystallites' hexagonal c-axes was confirmed by X-ray rocking
curve FWHM of $0.6^\circ$. Consequently, the anisotropy of the
resistivity is large with a ratio of the resistivities measured
along and normal to the c-axis of $5\times10^{3}$ at room
temperature.

\begin{figure}
\resizebox{\columnwidth}{!}{
  \includegraphics{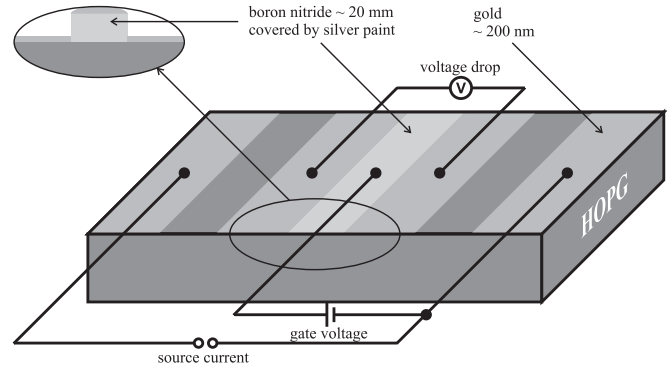}
} \caption{Scheme of the experimental setup. The vertical
dimensions in the enlarged frame are not to scale.} \label{setup}
\end{figure}

On top of the surface normal to the c-axis of the graphite lattice
four line contacts were made by sputtering of gold through a
shadow mask with a thickness of 200 nm (see Fig. \ref{setup}).
They were used to determine the resistance of the sample in a
four-point DC-measurement. The current has been inverted for each
point and the average of the resulting two voltages has been taken
to exclude thermoelectrical effects.

The main difficulty turned out to be the production of a thin but
yet stable dielectric layer. Many materials including
$\mbox{Al}_{2}\mbox{O}_{3}$ have been tested. Most of them did not
meet the experimental requirements, mainly due to the roughness of
the graphite surface and the poor adhesion between them an the
sample surface, which caused the layers to be porous and easy to
remove. A sprayed layer of boron nitride (BN) proofed to be stable
enough both from a mechanical and electrical point of view. A
solution of BN \cite{BfAM} received in a spray can was sprayed
between the two center gold contacts through a shadow mask and
then covered by silver paint as the gate electrode, because the
dielectric does not withstand the process of gold sputtering. A
tunable gate voltage $U$ between -100 V and +100 V was applied
between this electrode and the negative current electrode
(positive voltage corresponds to the negative pole connected to
the negative current electrode, see Fig. \ref{setup}).
Additionally, we have applied magnetic fields normal to the sample
surface.

As the BN-layer is mechanically soft, the standard methods for
determining the thickness fail. Therefore we estimated the
thickness by weighting the sample before and after the deposition
of the layer. The resulting thickness is $d = (20\pm10)
\mu\mbox{m}$. In the case of Si-MOSFET's the dielectric layer is
thinner by a factor of about 100, allowing the change of the
charge carrier density within a factor of 2 or more
\cite{Kravchenko}.

Assuming a charge carrier density in graphite of
$2\times10^{18}\mbox{cm}^{-3}$ \cite{Dresselhaus} and with the
spacing of the graphite layers $c/2 = 3.36\times10^{-8} \mbox{cm}$
one gets the carrier density of a single layer $n =
6.72\times10^{10} \mbox{cm}^{-2}$. In contrast, the effect of the
gate field can be estimated by $\Delta n \approx \epsilon_0U/d =
2.8\times10^{10} \mbox{cm}^{-2}$ at the maximum $U$. Taking into
account that probably more than one layer is active, the carrier
density changes by a factor below 0.4. Therefore the expected
effect is very small, if at all observable by our means.

\begin{figure}
\resizebox{\columnwidth}{!}{
  \includegraphics{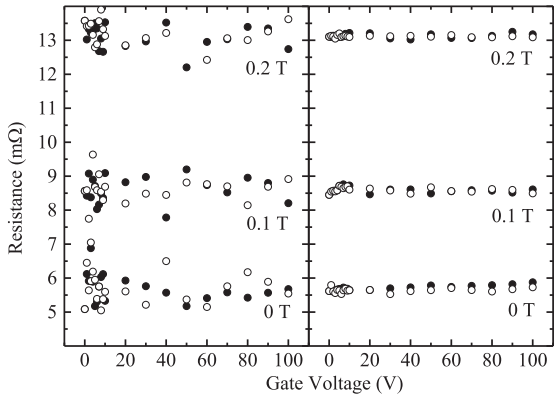}
} \caption{Resistance vs. gate voltage at a temperature of 2 K and
currents of 0.1 mA (left graph) and 1 mA (right graph) for
different magnetic fields (as labelled), positive ($\bullet$) and
negative ($\circ$) gate voltage.} \label{vdep}
\end{figure}

The dependence of the resistance on the gate voltage for two
different currents at different magnetic fields and at a
temperature of 2 K is shown in Fig. \ref{vdep}. Obviously there is
no visible effect of the gate field for none of the currents.
However, in the case of low current it could be masked by the
large scattering of the data.

\begin{figure}
\resizebox{\columnwidth}{!}{
  \includegraphics{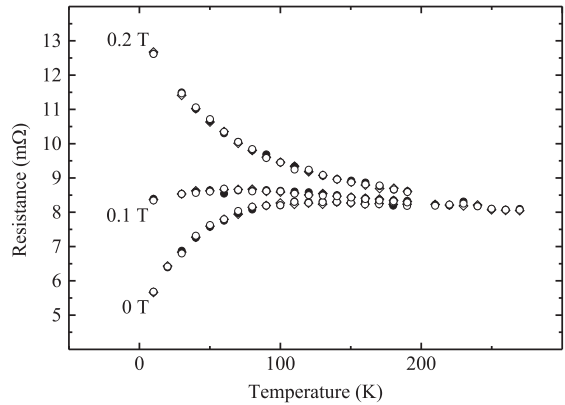}
} \caption{Resistance vs. temperature at a current of 1 mA and
different magnetic fields (as labelled). Different symbols
correspond to different gate voltages: $\diamond$ 0 V, $\bullet$
+100 V, $\circ$ -100 V.} \label{tdep}
\end{figure}

In Fig. \ref{tdep} we present the temperature dependence of the
resistance at different magnetic fields. The metal-insulator
transition is clearly seen. The resistance remains independent on
the gate field in all regimes.

By the arguments given above, the absence of an effect of the gate
field does not imply that there is none. It is highly desirable to
increase the range for changing the charge carrier density. A
favorable way to do so is the fabrication of a thinner insulating
layer serving as the gate dielectric in a field-effect transistor
configuration based on graphite. This remains a challenge for the
future.

\begin{acknowledgement}
We thank Y. Kopelevich for fruitful discussions and H. Hochmuth
for technical support. This work is supported by the Deutsche
Forschungsgemeinschaft under DFG ES 86/6-3.
\end{acknowledgement}


\begin{thebibliography}{}
\bibitem{Kempa1}
H. Kempa, Y. Kopelevich, F. Mrowka, A. Setzer, J.H.S. Torres, R.
H\"{o}hne, P. Esquinazi, Solid State Comm. {\textbf 115}, 539
(2000).
\bibitem{Kempa2} H. Kempa, P. Esquinazi, Y. Kopelevich, Phys. Rev.
B {\textbf 65}, 241101(R) (2002).
\bibitem{Kempa3} H. Kempa, H.C. Semmelhack, P. Esquinazi, Y. Kopelevich,
Solid State Comm. {\textbf 125}, 1 (2003).
\bibitem{Kopelevich1} Y. Kopelevich, P. Esquinazi, J.H.S. Torres,
R.R. da Silva, H. Kempa, F. Mrowka, R. Oca\~{n}a,\\
http://www.arxiv.org/abs/cond-mat/0209442.
\bibitem{Abrahams} E. Abrahams, S.V. Kravchenko, M.P. Sarachik,
Rev. Mod. Phys. {\textbf 73}, 251 (2001) and references therein.
\bibitem{Simonian} D. Simonian, S.V. Kravchenko, M.P. Sarachik,
V.M. Pudalov, Phys. Rev. Lett. {\textbf 79}, 2304 (1997).
\bibitem{Schoen1} J.H. Sch\"on, C. Kloc, B. Batlogg, Nature {\textbf 406}, 702 (2000).
\bibitem{Schoen2} J.H. Sch\"on, C. Kloc, B. Batlogg, Nature {\textbf 408}, 549 (2000).
\bibitem{Lee} J. Lee, G. Lientschnig, F. Wiertz, M. Struijk,
R.A.J. Janssen, R. Egberink, D.N. Reinhoudt, P. Hadley, C. Dekker,
Nano Lett. {\textbf 3}, 113 (2003).
\bibitem{Beasley} M.R. Beasley, S. Datta, H. Kogelnik, H. Kroemer, D.
Monroe,\\
http://www.lucent.com/news\_events/pdf/researchreview.pdf.
\bibitem{Ball} P. Ball, Nature {\textbf 421}, 878 (2003) and references
therein.
\bibitem{BfAM} B\"{u}ro f\"{u}r angewandte Mineralogie in
T\"{o}nisvorst, Germany.
\bibitem{Kravchenko} S.V. Kravchenko, W.E. Mason, G.E. Bowker,
J.E. Furneaux, V.M. Pudalov, M. D'Iorio, Phys. Rev. B {\textbf
51}, 7038 (1995).
\bibitem{Dresselhaus} M.S. Dresselhaus, G. Dresselhaus, Adv. Phys.
{\textbf 30}, 139 (1981).
\end{thebibliography}
\end{document}